\documentclass{article}
\usepackage{amsfonts}
\usepackage{amsmath} 
\usepackage{amsthm}
\usepackage{amssymb}
\usepackage{graphicx}
\usepackage{listings}
\usepackage{amsthm}
\usepackage[utf8]{inputenc}
\usepackage[T1]{fontenc}
\usepackage{lmodern}
\providecommand{\keywords}[1]{\textbf{\textit{Keywords---}} #1}
\newtheorem{lemma}{Lemma}[section]
\usepackage{subfig}
\title{Transactional Interpretation for the Principle of Minimum Fisher Information}
\date{November 5\,,2021}
\author{\textbf{Marcin Makowski\thanks{m.makowski@uwb.edu.pl},\, Edward W. Piotrowski}\\  Department of Mathematical Methods in Physics, \\University of Białystok
\and\textbf{Piotr Fr\k{a}ckiewicz} \\ Institute of Exact and Technical Sciences,\\ Pomeranian University in Słupsk
\and \textbf{Marek Szopa}\\Department of Operations Research,\\University of Economics in Katowice}

\begin{document}
\maketitle
\begin{abstract}The principle of minimum Fisher information states that  in the set of acceptable probability distributions characterizing the given system,  it is best done by the one that minimizes the corresponding Fisher information. This principle can be applied to transaction processes, the dynamics of which can be interpreted as the market tendency to minimize the information revealed about itself. More information involves higher costs (information is physical). The starting point for our considerations is a  description of the market derived from the assumption of minimum Fisher information for a strategy with a fixed financial risk.
Strategies of this type that minimize Fisher information overlap with the well-known eigenstates of a the quantum harmonic oscillator. 
The analytical extension of this field of strategy to the complex vector space (traditional for quantum mechanics) suggests the study of the interference of the oscillator eigenstates in terms of their minimization of Fisher information. 
It is revealed that the minimum value of Fisher information of the superposition of the \mbox{two strategies} being the ground state and the second excited state of the oscillator, has Fisher information less than the ground state of the oscillator. Similarly, less information is obtained for the system of strategies (the oscillator eigenstates) randomized by the Gibbs distribution. 
We distinguish two different views on the description of Fisher information. One of them, the classical, is based on the value of Fisher information. The second, we call it transactional, expresses Fisher information from the perspective of the constant risk of market strategies.  The orders of the market strategies derived from these two descriptions are different. From a market standpoint, minimizing Fisher information is equivalent to minimizing risk.\end{abstract}
\section{Introduction}
The relationship between the quantity of a good that producers want to sell at different prices and the quantity that consumers want to buy is a fundamental problem of \mbox{economics \cite{blaug}}. The relationships between supply and demand describe the process of market price formation. They are a source of information about market changes and their rate of change. We can illustrate this with supply and demand curves, which are graphical representations of the relationship between a product’s price and quantity. These curves can describe both the entire market (market curves) and characterize its participants (subjective curves): Buyers and sellers. The supply curve shows quantities of a good that a seller is willing to sell at different prices while the demand curve shows the quantity of a good that a consumer will buy at a certain price. Adopting a particular attitude towards buying/selling a given good at certain prices involves choosing a particular shape of the demand/supply curve (choosing a market strategy).

Mathematically, supply and demand curves are represented by functions that determine, respectively, supply and demand quantities for a given good as functions of its price. There is an interesting probabilistic interpretation of these functions in probability density terms.
Let us denote by $x$ the logarithm of the price per unit of some good. From the seller’s point of view, $x$ is the value for which they can sell a unit of the good and the buyer can either accept or reject this offer. The situation is similar from the point of view of the buyer, who can be viewed as the seller of the value of $x$ (e.g., expressed in money) per unit of the good. A market transaction is the process of exchanging goods. In this sense, the seller of a certain good buys money. On the other hand, the buyer of a certain good sells money. The money is just another good (like gold). The seller and buyer situations are symmetrical in the market. Both market participants attempt to approximate the supply/demand curves for the good based on historical data. We can interpret this as the probability of accepting the offer available in the market. Hence, the supply/demand curves define two random variables: $x$, the logarithm of the purchase price, and $y$, the logarithm of the selling price; and the curves themselves are probability distributions of these random variables:
\begin{eqnarray}
CDF_s(x)&=&\int_{-\infty}^{x}f_1(p)\,dp \hspace{1.5cm} (\textrm{supply}) \,,\nonumber\\
CDF_d(y)&=&\int_{y}^{\infty}f_2(p)\,dp \hspace{1.5cm} (\textrm{demand}) \,,\nonumber
\end{eqnarray}
where $ f_1 $ and $ f_2 $ represent (generally different) probability density functions. Interesting results based on the probabilistic interpretation of the supply/demand curves are described in \cite{De1, gold}, one of them is the expression of exceptions to the classical laws of supply and demand in the so-called negative probabilities \cite{neg}. 

Quantum mechanics has changed the manner in which we view the world. It is a theory that requires the observer to consider the effect of observational methods on the outcome of the observation. Complete and objective information about the phenomenon being studied is impossible, which is a fundamental principle of nature.
In our article, we refer to the quantum approach to the description of market strategies (supply and demand curves), adopting the principle of Minimum Fisher Information.
It states that in the set of admissible probability distributions characterizing the system being described, the one that minimizes the corresponding Fisher information characterizes it best. This principle can be applied to transactional processing the rate of change of which can be interpreted as the market’s tendency to minimize the information disclosed about it. This approach leads to a Schrödinger-type equation for the harmonic oscillator. We identify the solutions $\psi(x)$ of this equation as the market strategies that determine the supply and demand curves.

This paper is laid out as follows. In Section \ref{sec2}, we recall the definition of Fisher information. For clarity in the results presented, in the next Section \ref{sec3}, we present the derivation of the mentioned Schrödinger-type equation contained in \cite{Sch}. 
Next, we determine the Fisher information of the eigenvector (market strategy) of the oscillator. Furthermore, we determine the minimum Fisher information of the superposition of market strategies that constitute the eigenvectors of the oscillator and we determine the Fisher information of the system of these strategies randomized by the Gibbs distribution. Finally, we review the order of market strategies according to the description of Fisher information. We distinguish between two such descriptions. One of them, the classical one, we call the physical image. The second one, derived from the assumption of constant strategy risk, we call the transactional image.
\section{Fisher Information} \label{sec2}
The Fisher information $I_{F_\theta} $ is a measure of the amount of information carried by the observable random variable $X$  about an unknown parameter $\theta $ of the probability distribution $f(x;\theta)$ modeling $X$. It is defined as:
$$I_{F_\theta} = E \left[\left. \left(\frac{\partial}{\partial\theta} \log f(X;\theta)\right)^2\right|\theta \right] = \int_{-\infty}^{\infty} \left(\frac{\partial}{\partial\theta} \log f(x;\theta)\right)^2 f(x; \theta)\,dx.$$

In 
 our article, we consider the one-dimensional case by referring to the special, but often used in all kinds of applications, translation families \cite{Frieb, Frieb2}, which satisfy the condition:
$$f(x; \theta)=f(x-\theta)\,.$$
In this case, using the identity:
$$-\frac{\partial f}{\partial\theta}=\frac{\partial f}{\partial x}\,,$$ 
we can write the definitions of Fisher information as follows:
\begin{equation}\label{fish}
	I_{F}=\int_{-\infty}^{\infty} f(x)\bigg(\frac{d}{d x}\ln f(x)\bigg)^2  dx \,.
\end{equation}
The importance of Fisher information as a measure of information carried by probability distributions is known and widely discussed. It plays a fundamental role in the theory of estimation, which is reflected in the Cramér--Rao inequality: 
$$I_{F_\theta}\cdot\mathrm{var} \left(\hat{\theta}\right) \geqslant 1\,. $$
This inequality provides a lower bound for the variance of an unbiased estimator $\hat{\theta}$ of a parameter $\theta$.
Fisher information is important in the design of an experiment \cite{Frieb}, Bayesian inference (Jeffrey's distribution), and in fields that use statistical methods such as geophysics, biology, economics, signal analysis, and many more.

The concept of Extreme Physical Information (EPI) proposed by B.~Roy Friden states that many fundamental scientific laws can be derived from Fisher information. This is controversial \cite{kryt}, but research directions based on extreme values of Fisher information remain attractive and promising \cite{wy1, wy2,Fisher3}.  
Fisher information minimization leads (under certain conditions) to second-order differential equations that are often related to the fundamental laws of physics. In particular, minimizing Fisher information leads to Schrödinger-type equations for probability amplitudes. This forms the starting point of our work.
\section{Schr\"odinger-Like Equation from the Principle of Minimum Fisher \mbox{Information (MFI)}} \label{sec3}
Let $f(x)$ be the probability distribution 
of random variables $x$ (the logarithm of buying price) with the mean value $m$ and the corresponding risk $r$: 
 \begin{equation}
\label{brzegi}
1\,=\,\int_{-\infty}^{\infty}{f(x)}dx\,, \,\,\,m=\int_{-\infty}^{\infty}{x\, f(x)}dx\,, \,\,\,r=\int_{-\infty}^{\infty}{(x-m)^2\,f(x)}dx\,.
\end{equation}
 It is convenient to use probability amplitudes, for calculation reasons.
Let us denote $f(x) =:\psi^2(x)$, where $\psi(x)$ is a real-valued wave function. Using (\ref{fish}) we get the following formula for Fisher information:
\begin{equation}
	I_{F}=4\int_{-\infty}^{\infty}[\psi^{\prime}(x)]^2  dx \,.
\end{equation}
Finding the $\psi(x)$ function for which $I_F$ with the conditions (\ref{brzegi}) takes the minimum value, comes down to finding the minimum of the functional:
\begin{equation}
	\int_{-\infty}^{\infty} F(\psi(x),\psi^{\prime}(x),x)\,dx = \int_{-\infty}^{\infty} 4[\psi^{\prime}(x)]^2dx -\int_{-\infty}^{\infty} (a\,+\,b\,x\,+\,c\,x^2)\psi^2(x)dx\,, \label{fun} 
\end{equation}
where $ a $, $ b $, $ c $ are Lagrange multipliers. The extreme values of the functional (\ref{fun}), satisfying the Euler--Lagrange equations:
\begin{equation}\label{roww}
	\frac{d}{dx} \left(\frac{\partial F}{\partial \psi^{\prime}}\right) - \frac{\partial F}{\partial \psi} = 0\,.
\end{equation}
By re-parameterization of the $a$, $b$, $c$ coefficients 
 ($a=8\,\varepsilon\,\mu-4\, x_0^2\,\mu^2,\,b=8\,x_0\, \mu^2,$\,\mbox{$c=-4\,\mu^2$}) and translation $x\mapsto x+x_0-m$, we have (more details can be found in \cite{Sch}):
\begin{equation}
\label{zyta}
-\frac{1}{2\mu}\frac{\partial^2 \psi}{\partial x^2}+ \frac{\mu}{2}(x-m)^2\,\psi\,=\,\varepsilon \,\psi\,.
\end{equation}
This is the popular Schr\"odinger-type 
equation for a quantum harmonic oscillator. This model is one of the most ubiquitous in physics, playing a key role in many fields such as quantum optics, solid state physics, and many others.
It should be emphasized that \mbox{Equation (\ref{zyta})} was obtained by using only the  definition of Fisher information and assumption that $f(x)=:\psi^2(x)$, where $\psi(x)$ is a real-valued wave function. No other assumptions were necessary.
It is worth mentioning that Equation (\ref{zyta}) is referred to as the risk balance equation. We can call it that because the Fourier transform (FT) of Equation (\ref{zyta}) lead to an equation of the same type:
\begin{align}
	-\frac{d^2\psi(x)}{dx^2}\xrightarrow{FT} y^2\,\psi(y) & \,,\,\,\,\,\,\,\,\,\,\,\, \,\,\,\,\,x^2\,\psi(x)\xrightarrow{FT} -\frac{d^2\psi(y)}{dy^2}\,,\nonumber
\end{align}
where $\psi(y):=FT\,\psi(x)$. The term $(x-m)^2\,=(x-\langle x \rangle)^2$ of Equation (\ref{zyta}) is called the risk operator (selling risk or supply risk) because its expectation value: 
\begin{align}
 \langle (x-\langle x \rangle)^2\rangle = \int_{-\infty}^{\infty}(x-m)^2\,\psi^2 (x)\, dx\nonumber
\end{align}
 corresponds to the variance of a random variable $x$. The operator $-\frac{\partial^2}{\partial x^2}$ is associated with $(x-\langle x \rangle)^2$ by Fourier transform. Any market adopting the minimal supply strategy $\psi_n(x)$ does not reveal additional information if it shows the demand  $\psi_n (y):=FT\,\psi_n (x)$. The strategy that minimizes Fisher information on market is the same from the point of view of seller and buyer in the sense that they are mutually their Fourier transform. 
In the demand representation, $-\frac{\partial^2}{\partial x^2}$ takes the form $FT\, (-\frac{\partial^2}{\partial x^2}) \,FT=(y-\langle y \rangle)^2 $. This allows for its interpretation as the demand risk operator  and we have a connection between risk and information associated  with strategy: Minimal information content of a market supply/demand strategy is equal to the sum of the related supply and demand risk. Full details of the economic interpretation of this equation can be found in \cite{Sch}.
 \section{Fisher Information of Pure Oscillator State}
 The inequality:
\begin{equation}
8\,=\,\frac{\partial^2 F}{\partial {\psi^{\prime}}^2}\, >0\,,\nonumber
\end{equation}
implies that solutions of Equation (\ref{zyta}) are the minima of the functional (\ref{fun}) (see \cite{Fisher3}).
They form a discrete set of functions:
\begin{equation}
\psi_n(x)=\sqrt{\frac{\sqrt{\mu}}{2^n n! \sqrt{\pi}}}\,e^{-\frac{\mu\,(x-m)^2}{2}}\,H_n(\sqrt{\mu}(x-m)),\label{wlasny}
\end{equation}
$\varepsilon=\varepsilon_n=n+\frac{1}{2}$, for $n=0,1,2,\ldots$\, and $H_n(x)$ is  the $n$--th Hermite polynomial. 

This complete orthonormal set of functions stretches the vector space $L^2$ over the field $\mathbb{C}$ of square--integrable functions. The remaining solutions of Equation  (\ref{zyta})  (for negative values of $ \varepsilon $) are not square--integrable function (see \cite{pono}), so they cannot be used to construct a probability density function.

Let us denote by $I_{Fn}$, the Fisher information of the $n$--th eigenstate of the oscillator (\ref{wlasny}):
     $$
	I_{Fn}=4\int_{-\infty}^{\infty} \left(\frac{d\psi_n(x)}{dx}\right)^2dx \,. 
$$
Applying identity $\frac{dH_n(x)}{dx}=2nH_{n-1}(x)$ and recursive formula: $$H_{n+1}(x)=2xH_n(x)-2nH_{n-1}(x) $$  we obtain respectively:
\begin{equation}
\frac{d\psi_n(x)}{dx}=-\mu (x-m)\psi_n(x)+\sqrt{2n\mu}\,\psi_{n-1}(x)\,,\label{r4}
\end{equation}
\begin{equation}
\mu (x-m)\psi_n(x)=\sqrt{\frac{(n+1)\mu}{2}}\psi_{n+1}(x)+\sqrt{\frac{n\mu}{2}}\psi_{n-1}(x)\,.\label{r5}
\end{equation}
Substituting (\ref{r5}) into (\ref{r4}), we can rewrite the formula of the derivative of the function  $\psi_n(x)$:
\begin{equation}
\frac{d\psi_n(x)}{dx}=- \sqrt{\frac{(n+1)\mu}{2}}\psi_{n+1}(x)+\sqrt{\frac{n\mu}{2}}\psi_{n-1}(x)\,.\label{poch}
\end{equation}
Since $\int_{-\infty}^{\infty}\psi_m(x)\psi_n(x)dx=\delta_{nm}$, we see that:
\begin{equation}
I_{Fn}= 4\int_{-\infty}^{\infty} \left(\frac{d\psi_n(x)}{dx}\right)^2dx=(4n+2)\mu\,.\label{fisher}
\end{equation}
hence $I_{Fn}=4\mu\varepsilon_n$.  
\section{Fisher Information of Superposition of States with Locally Minimal $I_F$}
For the convenience of further calculations, we determine the values of two integrals.
 
\begin{lemma}
 Let $\psi_n$, $n=0,1,2,\ldots$ be the functions defined in (\ref{wlasny}), then:
\begin{align}
\nonumber v_{jk}:=&\int_{-\infty}^{\infty}(x-m)^2\psi_j(x)\psi_k(x)dx\\
=&\,\frac{1}{2\mu}\left(2\delta_{jk}\varepsilon_j 
+(\delta_{j+2,k}+ \delta_{j,k+2})\min_{r=j,k}\sqrt{(r+1)(r+2)} \right)\,,
\label{peku}\\
\nonumber w_{jk}:=&\,4\int_{-\infty}^{\infty}\frac{d\psi_j(x)}{dx}\frac{d\psi_k(x)}{dx}dx\\
=& \,2\mu\left(2\delta_{jk}\varepsilon_j 
-(\delta_{j+2,k}+ \delta_{j,k+2})\min_{r=j,k}\sqrt{(r+1)(r+2)} \right)\,.
\label{reku}
\end{align}
\end{lemma}
The Equalities (\ref{peku}) and 
(\ref{reku}) result from recursion   (\ref{r5}) and (\ref{poch}), respectively.  We must also remember about the orthonormality of states $\int_{-\infty}^{\infty}\psi_m(x)\psi_n(x)\,dx=\delta_{nm}$\,.

Let us consider the superposition of two eigenvectors $\psi_0$ and $\psi_k$:
\begin{equation}
\label{kok}
\psi_{p\alpha}:=\sqrt{p}\,\psi_0+e^{i\alpha}\sqrt{1-p}\,\psi_k\,,
\end{equation}
where $ p $ is the probability that $\psi_ {p\alpha}$ is in state $ \psi_0$ and $\alpha$
is the relative phase between $\psi_0$ and $\psi_k $
in this superposition.
The vector space corresponding to this superposition is
$\mathbb{C}^2$.

\begin{figure}[!h]
\includegraphics[width=9cm ]{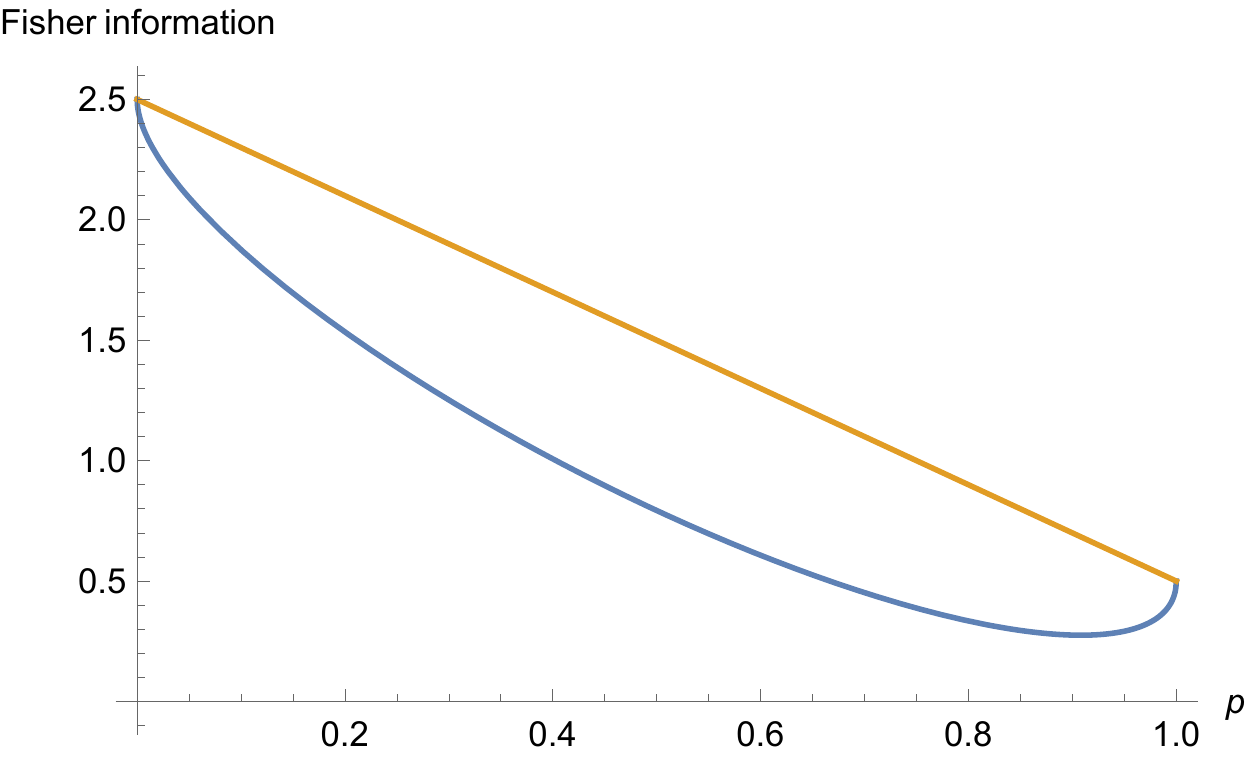}\caption{Graphs of functions $I_F(p,0)$ (lower) and $I_F(p,\pi/2)$ (upper, no interference effects) in units  $4\mu$\,.} \label{wyk}
\end{figure}

The Fisher information $(\ref{fish})$ extended to the domain of complex wave functions is given by:
\begin{equation}
\label{koko}
I_F=4\int \overline{\psi}^{\prime}_{p\alpha}(x)\,\psi'_{p\alpha}(x)~dx\,.
\end{equation}
Substituting  $(\ref{kok})$ into  $(\ref{koko})$ and applying $(\ref{reku})$ we obtain:
\begin{equation}\label{fsup}
I_F(p,\alpha):=p\,I_{F0}+ (1-p)\,I_{Fk}-4\mu \,\delta_{k,2}\cos(\alpha)\sqrt{2\,p\,(1-p)}\,.
\end{equation}
The smallest value on the right side (less than $ I_{F0} $) is obtained for $\alpha = 0 $ (interfering states have compatible phases) and $ p = p_{\min}: = \frac{1}{2} +6 ^ {-\frac{1}{2}} $ and $k = 2$. 
This value is:
\begin{equation}\label{minF}
I_F(p_{min},0)=(6-2\sqrt{6})\,\mu\approx 1.10102\,\mu.
\end{equation}
Figure \ref{wyk} shows the situation described here. Note that there are no interference effects between states $\psi_{0}$ and $\psi_{1}$.  This follows from Equation (\ref{fsup}) and  $\delta_{1,2}=0$.

To perform the physical process of transition from  $\psi_2 $ to $\psi_0$, which realizes the state--superposition with the least Fisher information we must have the appropriate Hamiltonian perturbation of the harmonic oscillator. Then $p$ is a monotone function of time during which this process of transition takes place. Systems realizing such situations (quantum games, information flow processes, quantum computers) must be based on appropriate quantum hardware (quantum game board), see e.g., \cite{Pio}.

The interference effects correct up to the second moment of superposition by the same amount by which the interference corrects down its Fisher information,  which is clear by comparing  Equations  (\ref{peku})  and  (\ref{reku}). 
The generalization of the above considerations into pairs of interfering states $\psi_k $ and $\psi_{k + 2} $ presents no additional difficulties.

At the end of this section, it is worth noting that we can look at Fisher information from a different point of view. That is, in the context of constant risk (fixed variance).  The variance of our superposition is:
$$
\mathrm{var} _{p_{\min},0}\left(\hat{x}\right)=\mu^{-1}(3/2-6^{-\frac{1}{2}})\,.
$$
Determining $\mu$ and substituting into (\ref{minF}) we obtain:
$$
I_F (p_{\min},0)= (11-4\sqrt{6})/\mathrm{var} _{p_{\min},0}\left(\hat{x}\right)\,.
$$
Since $11-4\sqrt{6}=1.2020\ldots> 1$, it follows that the Fisher information of our superposition is greater than the Fisher information (calculated in this way) of the ground state.  We will come back to such considerations in Section \ref{sec7}.

  \section{The Pure States of Harmonic Oscillator Randomized by Gibbs Distribution}
  Let us determine the probability distribution for the discrete family of functions $\psi_n(x)$, $n=0,1,2,\ldots$ that maximizes the Shannon entropy on the condition that the average value  of Fisher information is fixed. This type of analysis is successfully used in economics \cite{eco}.
  
  Let $ p_n $ be the probability of state $\psi_n(x)$. We wish to find the maximum of the function: 
$$-\sum_{n=0}^{\infty} p_n\ln p_n\,,$$ 
subject to the condition:
$$ \sum_{n=0}^{\infty} p_nI_{Fn} = \langle I_F\rangle, \,\,\,\,\,\,\,\, \sum_{n=0}^{\infty} p_n=1\,.$$
Using the Lagrange multipliers method, the problem of finding a distribution that maximizes the Shannon entropy leads to finding solutions of the equation (for each $p_n $):
\begin{equation} 
0=\frac{\partial}{\partial p_n}\bigg(-\sum_{n=0}^{\infty} p_n\ln p_n-\beta(\sum_{n=0}^{\infty} p_n I_{Fn} - \langle I_F\rangle) -\gamma(\sum_{n=0}^{\infty} p_n-1)\bigg)\,,\label{lag}
\end{equation}
where $\beta$, $\gamma$ are Lagrange multipliers.
The solutions of Equation (\ref{lag}) have the \mbox{following form}:
$$0=-\ln p_n -1-\beta I_{Fn} - \gamma .$$ 
Hence,
$$ p_n=e ^{-\gamma_0-\beta I_{Fn}}\,,$$
where $\gamma_0=1+\gamma$.
We have:
$$ p_n=\frac{e^{-\beta I_{Fn}}}{\sum_{k=0}^{\infty} e^{-\beta I_{Fk}}}\,,$$
because $\sum_{n=0}^{\infty} p_n=1$. It is a geometric distribution.
Applying (\ref{fisher}) we obtain:
\begin{equation} p_n=(1-e^{-d})\, e^{-d\, n}\, \label{geo}\end{equation}
where $d:=4\beta \mu >0$.
The probability density function $f_G$ for the set of eigenstates of the oscillator randomized with  distribution (\ref{geo}) is:
$$
f_G (x):= \sum_{n=0}^\infty p_n \psi^2_n(x)=\sqrt{\frac{\mu}{\pi}}
(1-e^{-d})e^{-\mu(x-m)^2}\sum_{n=0}^\infty \frac{1}{2^n\cdot n!} H^2_n(\sqrt{\mu}(x-m))e^{-d\, n}.
$$
The generating function of  square of Hermite polynomials has the following form \cite{herm}:
\begin{equation}
\label{twor}
\sum_{n=0}^{\infty}\frac{H_n^2(x)}{2^nn!}t^n=\frac{1}{\sqrt{1-t^2}}\exp\frac{2x^2t}{1+t}\,.
\end{equation}
Therefore,
$$
f_G(x)=\sqrt{\frac{s\mu }{\pi}}e^{-s\mu (x-m)^2}\,,
$$
where $s:=\frac{1-e^{-d}}{1+e^{-d}}\in(0,1)$. 
We can now use Formula (\ref {fish}) to find the Fisher information \mbox{of $f_G (x)$}:
\begin{equation}\label{fgib}
\langle I_F\rangle=:I_{FG}=2s\mu \,. 
\end{equation}

Since $s\in(0,1)$, the value of $I_{FG}$ is lower than Fisher information of the ground state $\psi_0$. In this case, the second moment is equal to:
\begin{equation}\label{rgib}
 r_G=\frac{1}{2s\mu}.
\end{equation}  

\section{Physical Order vs.~Transactional Order for Fisher Information} \label{sec7}

In Section \ref{sec3}, we determined the wave functions corresponding to the local minima of Fisher information using the method of Lagrange multipliers.
Their general form $(\ref{wlasny})$ depends on the parameter $\mu$. In any physical system that could implement market supply and demand strategies, this parameter is fixed and determined by such system-oscillator constants as its mass or the frequency of its oscillations. Let us call such a description of Fisher information its physical image 
and the corresponding order of values of Fisher information for its various local minima, physical order. In this order, Fisher information of the ground state of the oscillator closes the Fisher information domain $I_{FG}$ of the thermal states of the quantum oscillator from above, but
the smallest Fisher information value $I_F(p_{min},0)$ of the superposition of the oscillator’s eigenstates
is smaller than $I_{F0}$ and lies inside this domain. The corresponding values of Fisher information are:
$$I_{FG}=2s\mu,\,\,\,\,I_F(p_{min},0)=(6-2\sqrt{6})\mu, \,\,\,\, I_{Fn}=4\mu\varepsilon_n .$$ 
When we refer to market trading strategies, it is customary to characterize them by risk $r$ (variance of expected profit). Let us call this description of the strategy a transactional image. The inspiration comes from the Transactional Interpretation of Quantum \mbox{Mechanics \cite{cramer}}. To find the transactional order, $I_F$ corresponding to this view, one must represent the strategies in the parameter space of the system on the constraint (on the hypersurface) of the fixed value $r$ of the second moment of the system (risk). To do this, we need to take the multiplier $\mu$ out of the relevant formulas by equating the constraint on this second moment (the last Equation $(\ref{brzegi})$).

From now on we make the assumption: $r=const.$ In the case of the strategy being the  $n$th excited state of a quantum harmonic oscillator, the minimum Fisher information can be written as follows:
\begin{align}\label{r1}
	I_{Fn}= \frac{4}{r} \varepsilon_n^2\,.
\end{align}
 This follows directly from the definition of $r$ (\ref{brzegi}) and (\ref{peku}) for $j~=~k~=~n$ ($v_{nn}=r_n=r$). 

The risk $r$ of a strategy being the superposition of $\psi_0$ and $\psi_k$ states is given by the formula (we used $(\ref{peku})$):
\begin{align}
r&=\int_{-\infty}^{\infty}(x-m)^2\,\overline{\psi}_{p\alpha}\psi_{p\alpha} \,dx \\\nonumber
&=\frac{1}{\mu}\left(p\,\varepsilon_0+(1-p)\varepsilon_k+\cos(\alpha)\sqrt{2\,p\,(1-p)}\right)\,.
\end{align} 
When we substitute the determined $\mu$ into (\ref{fsup}) with $k = 2$ , we get: 

\begin{equation}\label{fsupG}
I_F(p,\alpha)=\frac{4}{r} \left((p\,\varepsilon_0+(1-p) \varepsilon_2)^2-2\cos^2(\alpha)\,p\,(1-p)\right)\,.
\end{equation}
The expression (\ref{fsupG}) takes the minimum for $p~=~p_{\min}~=~1$ and $\alpha \in [0,2\pi]$. We have:
\begin{equation}\label{2r}
I_F(p_{min},\alpha)=I_{F0}=\frac{1}{r}\,.
\end{equation}

Determining $\mu$ from  (\ref{rgib}) and substituting  into (\ref{fgib}), we obtain the relationship between minimum Fisher  information $I_ {FG}$ of eigenstates of quantum harmonic oscillator randomized by the Gibbs distribution and the risk $r=const.$
\begin{equation} \label{3r}
I_{FG}=\frac{1}{r}\,.
\end{equation}

\section{Conclusions}

The extension of the solutions of Equation (\ref{zyta}) to the linear complex space that stretches over them does not bring anything new to the eigenvalue problem.
However, it allows one to study Fisher information for any states-elements of this space, which are implemented in so many quantum physical systems. Such gameboards can be hardware for quantum markets. The information properties of the strategies used there are worth researching. The unique confidentiality features of quantum processes can encourage exploration of the possibilities of such markets. In our considerations, we assumed that any strategy-superposition is feasible in the quantum market. Such a market is an appropriate physical laboratory, i.e. a specific quantum gameboard (e.g., a quantum computer), where all possible unitary manipulations of quantum strategies and quantum measurements determining price conditions are possible (within a given set of qubits). Thus, the results of these measurements are classical information.

As is evident from the projective approach to the market description \cite{rzut}, the relevant invariant of this theory is not the profit from a single transaction but the profit from the corresponding buy-sell cycle. The same is true for the second moment of the variable $x$ (profit). To accurately assess the quality of a strategy, we should consider the aggregate risk of the buy--sell cycle executed with this strategy. The operators $-\frac{\partial^2}{\partial x^2}$ and  $(x-m)^2$ are their equivalents relative to the Fourier transform. Thus, we can choose the average energy of the system or the product (that is independent of $\mu$) of the expected values of these operators as a measure of risk over the entire cycle. The MFI principle in the transactional image leads to no difference between the Fisher information determined for the ground state of the harmonic oscillator (Formula (\ref{r1}) for $n=0$), the superposition of states, or any thermal states. For the latter, the Boltzmann--Shannon entropy is different from zero and different for different $\beta$.

In the suggested quantum model of market transactions, the player's strategy may (depending on the whim of the market) be read either in terms of supply or demand.
The Fisher information is nothing more than a Fourier image of the variance of the logarithm price, so it should be interpreted as the risk of an appropriate complementary market operation in the full buy/sell cycle, because only then is it possible to close the market position, i.e., to determine the profit.
Comparing the value of Fisher information for states with equal variance revealed that  the superposition $\psi_{p_{\min}0}$ has more information than the ground state $\psi_0$.
Only when the risks of supply strategies are the same, can one compare the information content of different demand strategies (and analogously for the \mbox{demand strategies}).

The analysis presented in this article showed an important relationship between Fisher information and risk. It could already be seen in the context of the Cramer--Rao inequality, which is so similar to the uncertainty principle in quantum mechanics.

\end{document}